# Experimental demonstration of quantum pigeonhole paradox

Ming-Cheng Chen[1,2] [*], Chang Liu[1,2] [*], Yi-Han Luo[1,2], He-Liang Huang[1,2], Bi-Ying Wang[1,2], Xi-Lin Wang[1,2], Li Li[1,2], Nai-Le Liu[1,2], Chao-Yang Lu[1,2], and Jian-Wei Pan[1,2]

[1] *Shanghai Branch, National Laboratory for Physical Sciences at Microscale and Department of Modern Physics, University of Science and Technology of China, Shanghai 201315, China;*
[2] *CAS Center for Excellence in Quantum Information and Quantum Physics, University of Science and Science and Technology of China, Hefei, Anhui 230026, China*

**Abstract**

We experimentally demonstrate that when three single photons transmit through two polarization channels, in a well-defined pre- and postselected ensemble, there are no two photons in the same polarization channel by weak-strength measurement, a counter-intuitive quantum counting effect called quantum pigeonhole paradox. We further show that this effect breaks down in second-order measurement. These results indicate the existence of quantum pigeonhole paradox and its operating regime.

> **Significance**
>
> We have demonstrated quantum pigeonhole paradox with three single photons. The effect of variable-strength quantum measurement is experimentally analyzed order by order and a transition of violation of the pigeonhole principle is observed. We find that the different kinds of measurement-induced entanglement are responsible for the photons' abnormal collective behavior in the paradox. The experimental violation of pigeonhole principle presents a challenge to the fundamental counting principle of Nature.

Quantum paradoxes (*1*) are at the frontier in the understanding of the fundamental quantum mechanics and have uncovered many strong conceptual conflicts transitioning from quantum to classical physics. Prominent examples include the Einstein-Podolsky-

Rosen (EPR) paradox that has introduced new physical concept such as nonlocality (*2,3*), and the quantum teleportation paradox that has brought numerous applications in quantum information sciences (*4,5,6*).

The pigeonhole principle (*7*), also known as the drawer principle or shelf principle, states that if there are more pigeons than boxes, at least one box must contain more than one pigeon. Recently, Aharonov *et al.* (*8*) put forward a quantum pigeonhole paradox where the classical pigeonhole counting principle in some case may break down. It was shown that three two-state quantum particles pre- and postselected in a particular subensemble could result in no two particles being in the same quantum states.

To observe the quantum pigeonhole effect, an elaborated quantum measurement is required to probe the single quantum particles. Quantum measurement (*9-11*) that bridges the quantum and classical world has played a key role in many counter-intuitive quantum protocols, ranging from quantum teleportation (*4,12,13*) to interaction-free measurement (*14,15*). While projection measurements are routinely used to estimate the expectation of physical observables, they inevitably disturb or destroy the studied physical system. To interrogate the states of quantum particles but with minimal measurement-induced disturbance, the weak-strength measurement invented by Aharonov *et al.* ~30 years ago provides a feasible way *(16)*.

Here we adopt the weak measurement concept to probe the underlying physics of quantum pigeonhole paradox. In our experiment, we use three quantum pigeons, which are three single photons, and two "boxes", which are denoted by the two orthogonal polarization states horizontal (*H*) and vertical (*V*). The photons are initially prepared in the state $|\psi\rangle = |+\rangle_1 |+\rangle_2 |+\rangle_3$ and finally measured and post-selected in circular polarization state $|\phi\rangle = |R\rangle_1 |R\rangle_2 |R\rangle_3$, where $|+\rangle = (|H\rangle + |V\rangle)/\sqrt{2}$, $|R\rangle = (|H\rangle + i|V\rangle)/\sqrt{2}$, and the subscripts labels the three photons. The success probability to observe the postselected state is $|\langle \psi | \phi \rangle|^2 = 1/8$. To probe whether two quantum pigeons are in the same box, we introduce parity check observables $S_{ij} = |H_i H_j\rangle\langle H_i H_j| + |V_i V_j\rangle\langle V_i V_j|$ to query whether the *i*-th and *j*-th photons are in the same polarization state but without measuring which polarization. Similarly, we use a three-body parity check observable $S_{123} = |H_1H_2H_3\rangle\langle H_1H_2H_3| + |V_1V_2V_3\rangle\langle V_1V_2V_3|$ to query whether the three photons are all in the same polarization. It is obvious that non-zero observable $S_{123}$ implies non-zero observables $S_{ij}$, but the converse is not true.

To observe the pre- and postselected quantum systems without disturbance, weak-strength measurement of the observables is required *(8)*. We introduce a quantum meter system with initial state $|\varphi\rangle$, which will be changed by non-zero observables $S_{ij}$ according to system-meter interacting Hamiltonian:

$$H = S_{12} \otimes F_m + S_{23} \otimes F_m + S_{31} \otimes F_m. \tag{1}$$

Here the meter state $|\varphi\rangle$ is a narrow wave packet which can be moved by the displacement operator $F$, as shown in Fig. 2(a). With an interaction strength $\lambda$, the meter is moved to a new quantum state

$$|\varphi'\rangle = \langle\phi|e^{-i\lambda H}|\psi\rangle|\varphi\rangle \tag{2}$$

We expand the exponential operator according to the interaction strength $\lambda$ up to the second order:

$$\begin{aligned}|\varphi'\rangle \approx &\langle\phi|\psi\rangle|\varphi\rangle - i\lambda(\langle\phi|S_{12} + S_{23} + S_{31}|\psi\rangle) \otimes F_m|\varphi\rangle \\ &- \frac{\lambda^2}{2}(\langle\phi|S_{12} + S_{23} + S_{31} + 6S_{123}|\psi\rangle) \otimes F_m^2|\varphi\rangle + O(\lambda^3)\cdots\end{aligned} \tag{3}$$

This formula shows that the shift of meter is proportional to $\langle\phi|S_{12} + S_{23} + S_{31}|\psi\rangle$ and $\langle\phi|S_{12} + S_{23} + S_{31} + 6S_{123}|\psi\rangle$ in the first and second order of $\lambda$, respectively.

For small $\lambda$, the system is almost not disturbed, and we expect to observe first-order meter shifts when two photons appear in the same polarization. If no first-order shift is observed, it implies that arbitrary two photons are not in the same polarization and proves the emergence of quantum pigeonhole paradox. In this first-order effect, the coefficients $\langle\phi|S_{ij}|\psi\rangle$ completely determine the meter shift, and coefficient $\langle\phi|S_{123}|\psi\rangle$ plays a non-negligible role in high-order effect.

In the past few years, weak measurement on single photon has been implemented in numerous experiments *(17-24)*. However, directly performing a weak measurement on three photons simultaneously is still out of current photonic quantum technology. Here we circumvented this challenge by directly analyzing the expected first- and second-order meter shift. Below we describe our designed photonic experiments to probe the parity observables of the pre- and postselected ensembles.

The experimental setup is shown in Fig. 2(b). Three polarization-encoded single photons are prepared from two pairs of entangled photons produced by spontaneous parametric down-conversion (SPDC) process *(25)*. The experiment constitutes three stages: (1) preparing the photons in the polarization state $|\psi\rangle$; (2) probing the photons polarization parities; (3) measuring the photons in circular polarization bases.

In the first stage, two beam-like type-II beta-barium borate (BBO) crystals are pumped by ultraviolet laser pulses with a central wavelength of 394 nm *(26)*. The emitted four photons are filtered by polarizing beam splitters (PBSs) and prepared by half-wave plates (HWPs) in the diagonal polarization states. The fidelity of the prepared polarization state of three photons is 0.992(4), as shown in Fig. 3(a). One extra photon is traced out and the other three photons are passed to the second stage.

In the second stage, PBSs are used to check the polarization parities. A PBS has two input and two output ports, and transmits photon's *H* component and reflects its *V* component. To ensure the two injected photons to emit from two output ports, the two photons must both transmit or both reflect simultaneously. That is when we register two photons in two output ports, we know that the photons have the same polarization components but still don't know which one *(27)*. In this sense, a PBS performs the quantum operator $S_{ij}$ on the two coming photons. Adding a second PBS, the three-body operator $S_{123} = S_{23}S_{12}$ can also be implemented. We performed two-photon Hong-Ou-Mandel (HOM) interference *(28,29)* to guarantee the photons arriving the PBS are of maximal indistinguishability. The HOM interference visibilities from same SPDC photons and different SPDC photons are 0.989(1) and 0.90(2), respectively, as shown in Fig. 3(b). The interference visibilities are stable during the experimental period and the reduction of interference visibility between different SPDC photons is due to the spectral correlation between the interference photon and its twins SPDC photon *(26)*.

In the final stage, the three single photons are measured in circular polarization bases by quarter-wave plates (QWPs) followed by PBSs. We recorded the three-fold coincident counts and showed in Fig. 4. In Fig. 4(a), a PBS is used to perform the quantum operation $S_{12}$ on photons 1 and 2. Totally 1105 events are recorded and there are only 3 events projected to the post-selected state $|\phi\rangle$. So, the expectation of parity observable $|\langle\phi|S_{12}|\psi\rangle|^2 / |S_{12}|\psi\rangle|^2$ is estimated to be 0.002(1), which is of significant zero probability. The small residual value above zero is mainly contributed by the imperfect multi-photon destructive interference, characterized by the finite HOM interference visibilities of two individual single photons. Furthermore, due to the apparent symmetry between the three photons in our test system, we can conclude that the parity of any pair of two photons is all zero. In Fig. 4(b), We show the counts of three-fold events using two PBSs to estimate the three-body parity observable. Here 1152 events are recorded, where 269 events are projected to the post-selected state $|\phi\rangle$. We estimate that $|\langle\phi|S_{123}|\psi\rangle|^2 / |S_{123}|\psi\rangle|^2 = 0.23(1)$, which is significantly non-zero.

Our experimental results show that in the defined pre- and postselected three photons subensemble, an imaginary non-invasive weak measurement will not observe a first-order meter shift due to the zero observables $|\langle\phi|S_{ij}|\psi\rangle|^2$ and indicates that no two photons are in the same polarization channel, proving the quantum pigeonhole paradox. When the measurement to probe the parity becomes stronger, the non-zero second-order meter shift (proportional to $|\langle\phi|S_{123}|\psi\rangle|^2$) is not negligible and we can observe that the probability of three photons in the same polarization is significantly non-zero, which naturally implies two photons could in the same box. Therefore, the phenomenon of quantum pigeonhole paradox can only be observed by non-invasive measurement and will disappear in strong measurement.

We note that the parity observations induce quantum correlations in the single photons. With small measurement strength, the measurement back-action of observables $S_{ij}$ produce an EPR-type entanglement *(2)*, while at stronger measurement, high-order back-action $S_{ij}S_{jk}$ produce a Greenberger-Horne-Zeilinger (GHZ) -type entanglement *(30)*. These induced quantum entanglements are responsible for the collective properties of independent quantum particles in quantum pigeonhole paradox, controlling the occurrence or disappearance of quantum pigeonhole effects.

In summary, we have implemented quantum pigeonhole paradox by transmitting three single photons through two polarization channels, which demonstrates pigeonhole paradox definitely at the level of individual quantum particles *(31)*. The experiments show that in addition to the condition of well-defined pre- and postselected subensemble, non-invasive weak measurement is also required to observe the counting paradox. We implement the desired measurement indirectly by analyzing the measurement effects order by order and reveal the paradox will not survive under high-order measurement.

**References and Notes**

1. Aharonov Y, Rohrlich D (2008) *Quantum paradoxes: quantum theory for the perplexed* (John Wiley & Sons).

2. Einstein A, Podolsky B, Rosen N (1935) Can quantum-mechanical description of physical reality be considered complete? *Physical review* 47(10):777.

3. Bell JS (2001) Einstein-Podolsky-Rosen experiments. John S Bell on the Foundations of Quantum Mechanics (World Scientific), pp 74–83.

4. Bennett CH, et al. (1993) Teleporting an unknown quantum state via dual classical and Einstein-Podolsky-Rosen channels. *Physical review letters* 70(13):1895.

5. Raussendorf R, Briegel HJ (2001) A one-way quantum computer. *Physical Review Letters* 86(22):5188.

6. Kimble HJ (2008) The quantum internet. *Nature* 453(7198):1023.

7. Allenby RB, Slomson A (2011) *How to count: An introduction to combinatorics*

(CRC Press).

8. Aharonov Y, *et al.* (2016) Quantum violation of the pigeonhole principle and the nature of quantum correlations. *Proceedings of the National Academy of Sciences* 113(3):532–535.

9. von Neumann J, Rose M (1955) Mathematical Foundations of Quantum Mechanics. *Physics Today* 8:21.

10. Schlosshauer M (2005) Decoherence, the measurement problem, and interpretations of quantum mechanics. *Reviews of Modern physics* 76(4):1267.

11. Wheeler JA, Zurek WH (2014) *Quantum theory and measurement* (Princeton University Press).

12. Bouwmeester D, *et al.* (1997) Experimental quantum teleportation. *Nature* 390(6660):575.

13. Pirandola S, Eisert J, Weedbrook C, Furusawa A, Braunstein SL (2015) Advances in quantum teleportation. *Nature photonics* 9(10):641.

14. Elitzur AC, Vaidman L (1993) Quantum mechanical interaction-free measurements. *Foundations of Physics* 23(7):987–997.

15. Kwiat P, Weinfurter H, Herzog T, Zeilinger A, Kasevich MA (1995) Interaction-free measurement. *Physical Review Letters* 74(24):4763.

16. Aharonov Y, Albert DZ, Vaidman L (1988) How the result of a measurement of a component of the spin of a spin-1/2 particle can turn out to be 100. *Physical review letters* 60(14):1351.

17. Pryde G, O'Brien J, White A, Ralph T, Wiseman H (2005) Measurement of quantum weak values of photon polarization. *Physical review letters* 94(22):220405.

18. Goggin ME, *et al.* (2011) Violation of the Leggett–Garg inequality with weak measurements of photons. *Proceedings of the National Academy of Sciences* 108(4):1256–1261.

19. Kocsis S, *et al.* (2011) Observing the average trajectories of single photons in a two-slit interferometer. *Science* 332(6034):1170–1173.

20. Lundeen JS, Sutherland B, Patel A, Stewart C, Bamber C (2011) Direct measurement of the quantum wavefunction. *Nature* 474(7350):188.

21. Rozema LA, *et al.* (2012) Violation of Heisenberg's measurement-disturbance relationship by weak measurements. *Physical review letters* 109(10):100404.

22. Salvail JZ, *et al.* (2013) Full characterization of polarization states of light via direct measurement. *Nature Photonics* 7(4):316–321.

23. Malik M, *et al.* (2014) Direct measurement of a 27-dimensional orbital-angular-momentum state vector. *Nature communications* 5:3115.

24. Thekkadath G, *et al.* (2016) Direct measurement of the density matrix of a quantum system. *Physical review letters* 117(12):120401.

25. Kwiat PG, *et al.* (1995) New high-intensity source of polarization-entangled photon pairs. *Physical Review Letters* 75(24):4337.

26. Wang X-L, *et al.* (2016) Experimental ten-photon entanglement. *Physical review letters* 117(21):210502.

27. Pan J, Zeilinger A (1998) Greenberger-horne-zeilinger-state analyzer. *Physical Review A* 57(3):2208.

28. Hong C-K, Ou Z-Y, Mandel L (1987) Measurement of subpicosecond time intervals between two photons by interference. *Physical review letters* 59(18):2044.

29. Pan J-W, Daniell M, Gasparoni S, Weihs G, Zeilinger A (2001) Experimental demonstration of four-photon entanglement and high-fidelity teleportation. *Physical Review Letters* 86(20):4435.

30. Greenberger DM, Horne MA, Shimony A, Zeilinger A (1990) Bell's theorem without inequalities. *American Journal of Physics* 58(12):1131–1143.

31. Anjusha V, Hegde SS, Mahesh T (2016) NMR investigation of the quantum pigeonhole effect. *Physics Letters A* 380(4):577–580.

**Figure captions**

Figure 1: Quantum pigeonhole paradox. (a) Transmitting three photons through two polarization channels. In the classical world, there must be at least two photons through the same channel. In the quantum world, this will be no longer always true. (b) The proposed experiment. Three photons are prepared and detected in certain polarization bases, and non-invasive quantum measurements are used to probe the photons polarization state.

Figure 2: The experimental scheme. (a) A strength-variable measure system. A virtual meter is used to probe the polarization parity of photons. (b) Experimental setups. Ultraviolet (UV) femtosecond laser pulses (394 nm, 80 MHz) pass through two beam-like type-II compound beta-barium borate (C-BBO) crystals to produce four single photons by spontaneous parametric down-conversion (SPDC) process. The photons are prepared in diagonal polarization product states with polarization beam splitters (PBSs) and half-wave plates (HWPs). The parity observations are implemented by overlapping two single photons on PBSs. Finally, the photons are detected in circular polarization bases with quarter-wave plates (QWPs), HWPs and PBSs, and the four-fold photon events are recorded.

Figure 3: The experimental calibrations. (a) The prepared diagonal polarization states. The Fidelity is 99.6%. (b) HOM interference of same SPDC photons. The visibility is 99%. (b) HOM interference of different SPDC photons. The visibility is 90.5%

Figure 4: The experimental results. (a) Probing two-photons parity. The probability of projection on to post-selected state $|RRR\rangle$ is significantly zero. (b) Probing three-photon parity. The probability of projection on to post-selected state $|RRR\rangle$ is significantly non-zero.

**Acknowledgments**: This work is supported by the National Natural Science Foundation of China, the Chinese Academy of Sciences, the National Fundamental Research Program, the Anhui Initiative in Quantum Information Technologies, and the Postdoctoral Innovation Talents Support Program. We thank Hongyi Zhou and Xiongfeng Ma for helpful discussions.

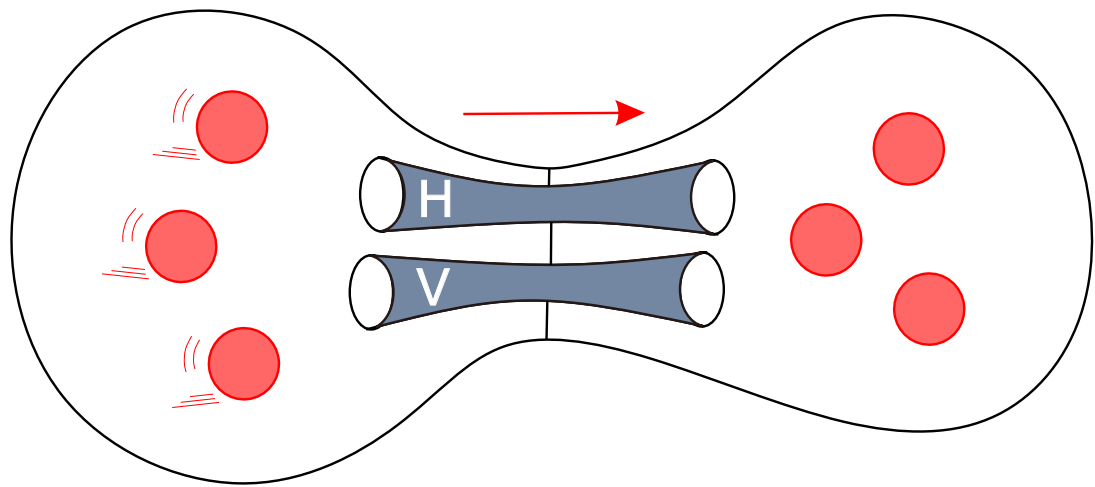

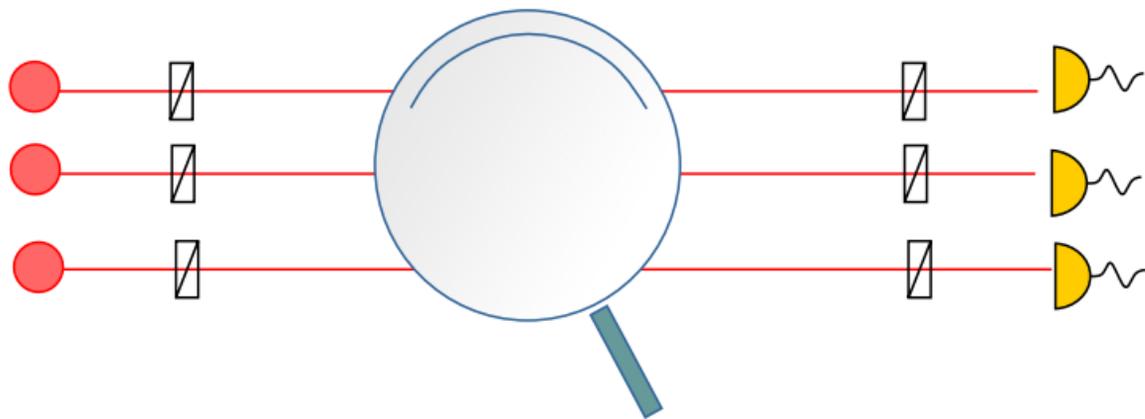

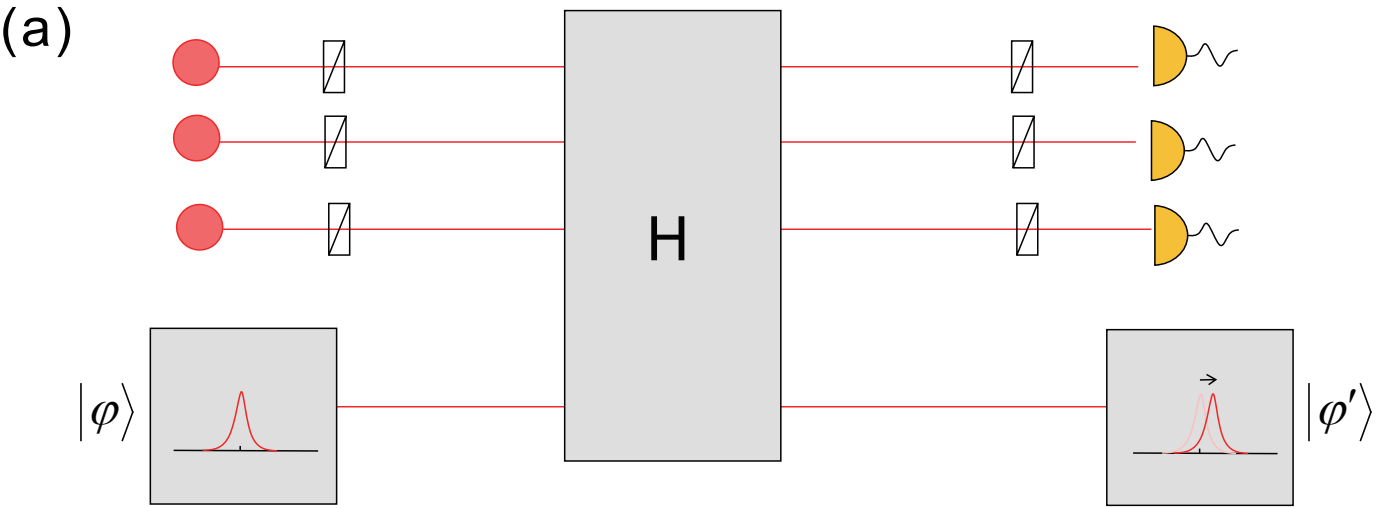

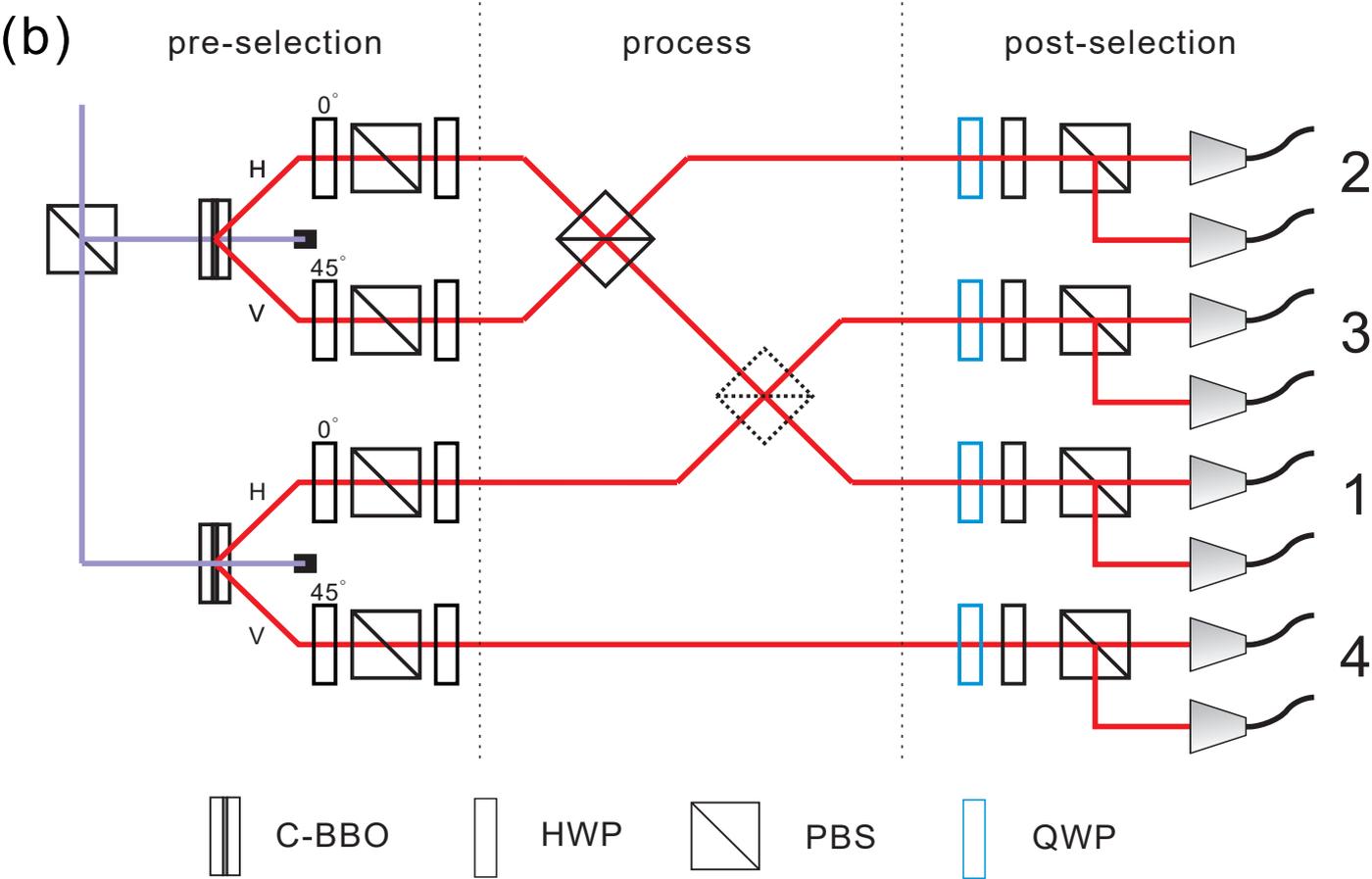

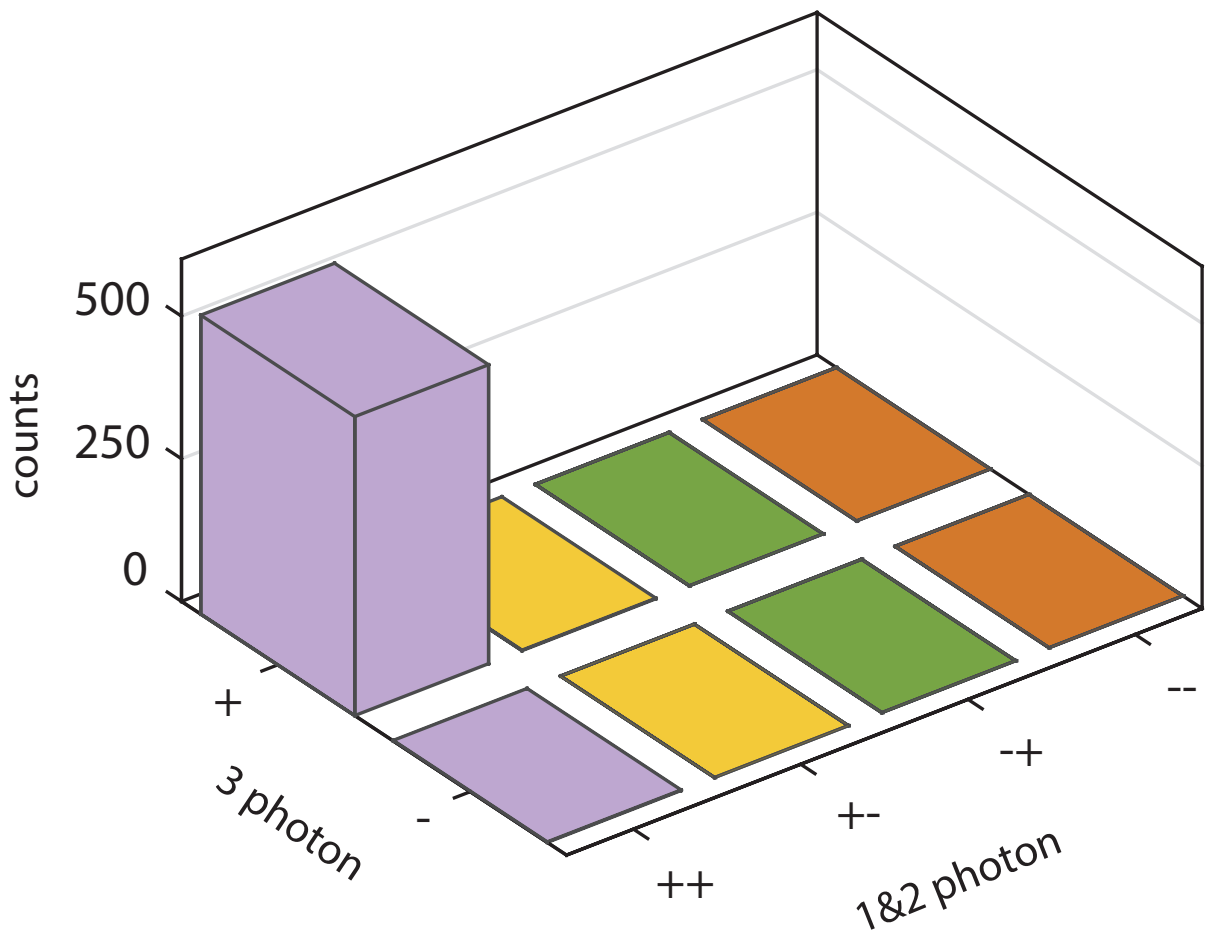 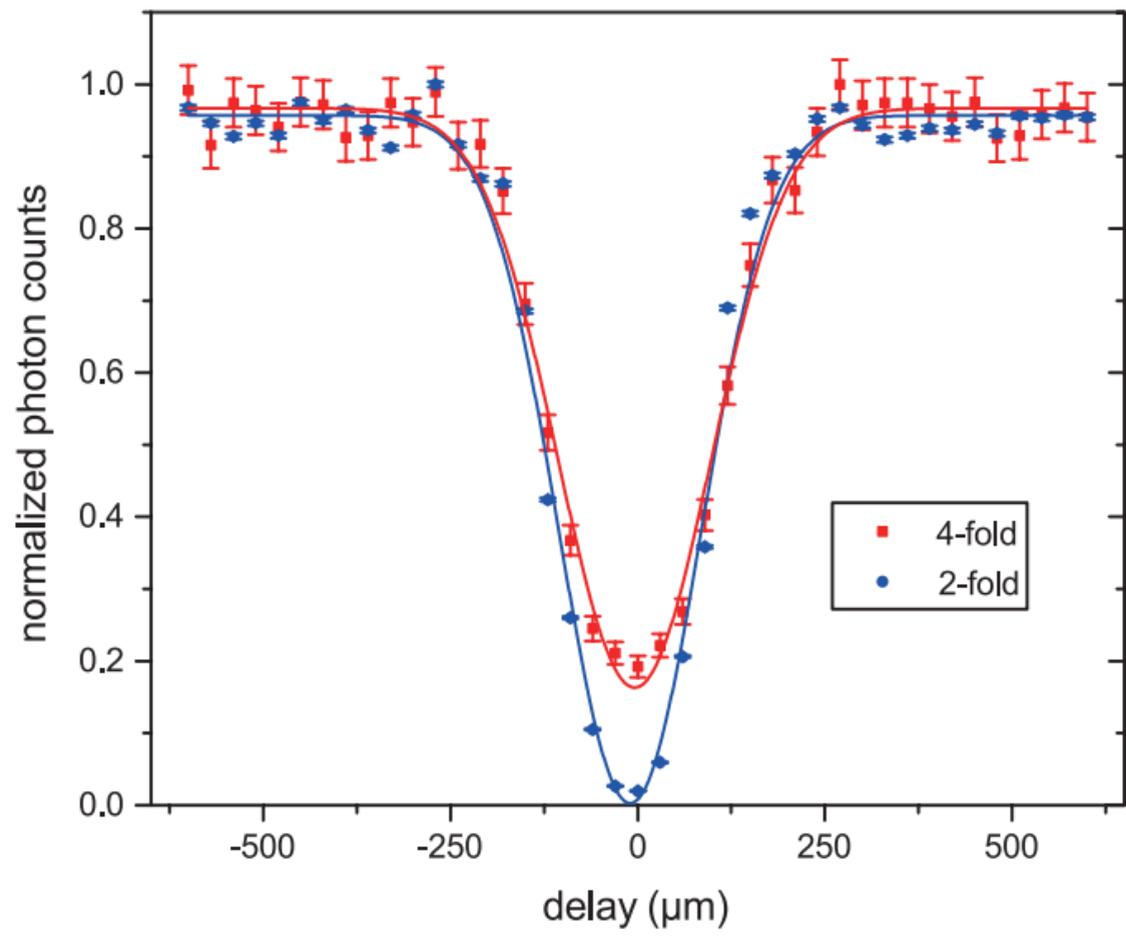

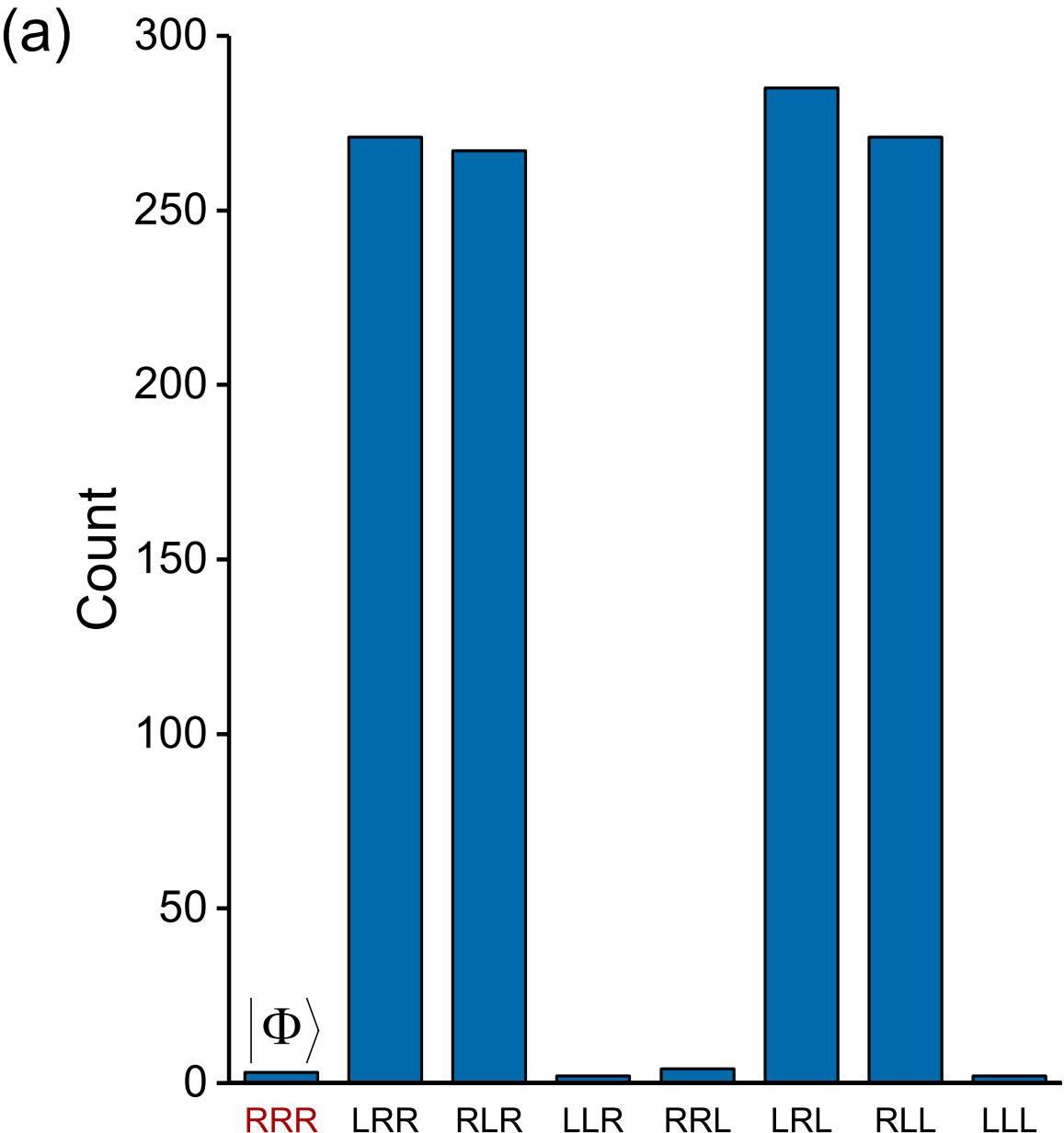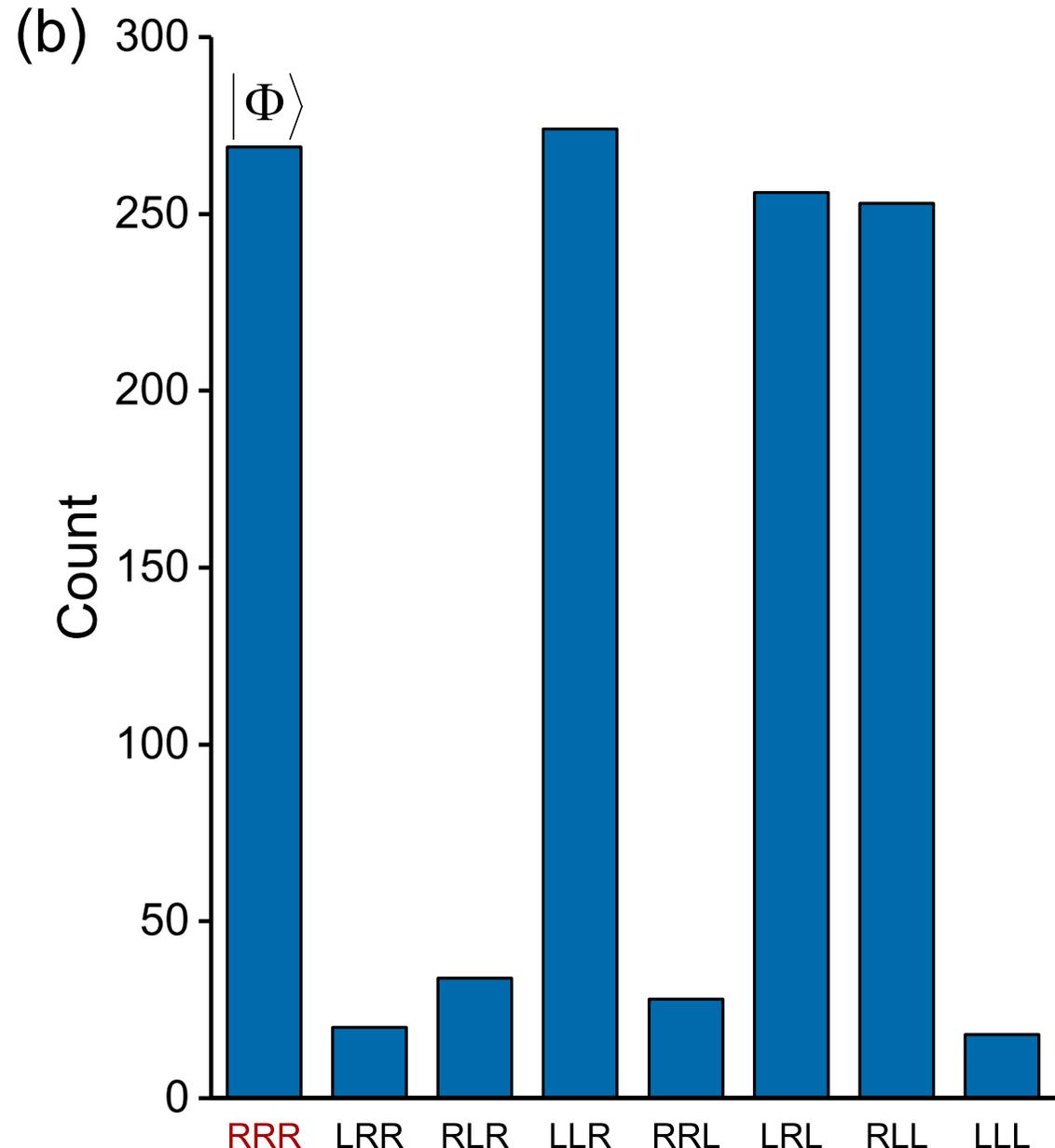